\documentstyle[prl,aps,epsf,floats]{revtex}

\def\ve{\varepsilon} \def\aup{\uparrow} \def\adw{\downarrow}

\def\be{\begin{equation}}  \def\ee{\end{equation}}
\def\bea{\begin{eqnarray}}  \def\eea{\end{eqnarray}}
\def\bs{\bigskip}

\begin{document}

\draft

\twocolumn[\hsize\textwidth\columnwidth\hsize\csname@twocolumnfalse%
\endcsname

\title{The Cooperon and the Random Matrix Model for Type-II Superconductors}
\author{Safi R. Bahcall}
\address{Department of Physics, University of California, 
Berkeley CA 94720}
\maketitle

\begin{abstract}

We derive the connection between the Cooperon problem in weak
localization theory and the random matrix description of type-II
superconductors.  As magnetic field and disorder increase, an extreme
type-II superconductor crosses over from a state in which the low
energy quasiparticles are primarily localized and the density of
states is determined by the electronic structure of individual
vortices, to a `chaotic' state, in which quasiparticles are primarily
extended and the density of states is determined by the random
matrix description.

\end{abstract}


\vspace*{10pt}


 ]
 \def\bottomfraction{.9}
 \def\textfraction{.1}

\section{Introduction}

In this paper we explore the hypothesis that electrons in a
superconductor with explicit time-reversal breaking pair via random
matrix elements.  This hypothesis combines ideas and techniques from
three fields of study usually considered unrelated: random matrices,
type-II superconductors, and the Cooperon problem of weak
localization.  We will show that incorporating these ideas into a
random matrix model allows us to calculate the density of states of a
type-II superconductor non-perturbatively and leads to new
possibilities for investigating chaotic behavior in complex quantum
systems.

This section contains brief introductions, with references to
extensive reviews, of the main ingredients of the theory.  In section
II we describe the random matrix model, a generalization of the
Anderson description of disordered superconductors to superconductors
in a magnetic field.  In section III we derive the connection between
this description of type-II superconductors and the Cooperon.  In
section IV we show that this connection yields the correct answer for
the semiclassical upper critical field, $H_{c2}$.  In section V we
discuss the weak field limit and the relation between the random
matrix model and the Abrikosov-Gor'kov theory.  Finally, in section VI
we summarize the principal results and describe open questions and
further possible areas of research.

\subsection{Random Matrices}

Random matrices were introduced into theoretical physics by Wigner
in the 1950's as a tool for understanding the distribution of energy
level spacings and widths observed in nuclei \cite{wigner}.  The
distribution of level spacings in nuclei was compared with the
distribution of level spacings of a large matrix filled with random
(uncorrelated) matrix elements; quantitatively good agreement was
eventually found \cite{nucleireview}.  The idea of comparing the level
spacings of complex quantum systems with the level spacings of random
matrices was eventually extended to atoms \cite{porter}, small
metallic particles \cite{RMandmetals}, and classically chaotic systems
\cite{RMandchaos}.  In all of these applications, in condensed matter,
atomic, and nuclear physics, the focus has been on understanding the
distribution of the {\it spacings} of energy levels, that is, the
probability $P(E,x)dx$ that the neighbor of a level at energy $E$ has
an energy in the range $E+x$ to $E+x+dx$.  In contrast, we will use
random matrices to understand the {\it density} of energy levels, that
is, the number of energy levels $\rho(E)dE$ in the range $E$ to
$E+dE$.

Random matrices have also been studied in the context of field theory,
spurred by 't Hooft and Brezin et al.'s discovery of the connection
with the large-N limit of certain SU(N)-invariant field theories
\cite{RMandft}.  A connection between the Feynmann diagram expansion
for random matrix theories and the sum over random surfaces in
two-dimensional quantum gravity was also established
\cite{RMandgravity}.  More recently, random matrices have resurfaced
as a possible unification of various string theories
\cite{RMandstrings}, and as a concrete example of a more general
theory of ``non-commuting random variables'' \cite{RMandmath}.

Extensive recent reviews can be found in Refs.\ \cite{mehta} and
\cite{guhr}.

\subsection{Type-II Superconductors}

The BCS theory \cite{BCS} describes the electronic properties of
ideal, uniform superconductors; it applies to the interiors of type-I
superconductors, which expel magnetic field, or to type-II
superconductors below the field of first penetration, $H_{c1}$.  The
presence of a magnetic field, which occurs for type-II superconductors
in the mixed state, $H_{c1}<H<H_{c2}$, makes the calculation of
electronic properties significantly more complicated.  The magnetic
field destroys both the spatial uniformity and the time-reversal
symmetry implicit in the BCS description.

Magnetic field penetrates a superconductor in the form of tubes of
magnetic flux (vortices) enclosing circulating quasiparticle currents.
These currents lead to electronic states which are bound to individual
vortices.  Such states are not only spatially inhomogeneous, but may
overlap strongly with neighboring vortices and form energy bands, the
details of which will be sensitive to the presence of disorder or
the deviations from an ideal vortex lattice that are found in most
materials.

It is possible to attack the problem of the electronic properties of
type-II superconductors numerically.  For a fixed type of vortex
lattice, a large unit cell can be considered which contains one or
more vortices.  Either the full Bogoliubov-de Gennes equations
\cite{degennes} or the approximate Eilenberger equations
\cite{eilenberger} (derived by assuming slow variations on the scale
of the Fermi wavelength) may be solved to obtain the quasiparticle
eigenstates \cite{norman,japanese}.  

The numerical approach to the electronic properties of type-II
superconductors has several drawbacks: (1) It is difficult to get
useful results numerically, both because of the large number of
electrons per unit cell for realistic materials and because of the
widely different energy scales in the problem.  (The cyclotron
frequency is much less than the BCS gap, which is in turn much less
than the Debye energy.)  (2) In a magnetic field, unlike conventional
superconductors at zero field, the electronic properties are sensitive
to disorder.  The highly idealized impurity-free perfect vortex
lattice is unrealistic for most materials.  (3) The numerical approach
calculates every quasiparticle wavefunction in the superconductor.
This is often an overkill: for many applications we need much less
information, for example, just the single-particle density of states.
An approach which is simpler and more general than numerically solving
the Bogoliubov-de Gennes equations is desirable.

Analytically manageable extensions of the BCS theory to disordered
superconductors were given both by Anderson \cite{anderson59} and by
Abrikosov and Gor'kov \cite{abrigork}.  Anderson generalized the BCS
description to show that non-magnetic impurities have little effect on
conventional superconductors.  Abrikosov and Gor'kov introduced a
diagrammatic technique to include the effects of a random ensemble of
impurities, and showed that magnetic impurities can suppress
superconductivity.

Neither the Anderson nor the Abrikosov-Gor'kov description, however,
applies directly to the mixed state of type-II superconductors. The
Abrikosov-Gor'kov description requires spatial homogeneity, which is
violated in the mixed state, and, in order to apply to a type-II
superconductor, must assume that random magnetic impurities are
equivalent to an applied magnetic field \cite{maki}.  The
Anderson description is more general, in the sense that it is not tied
to the assumption of averaging over a random ensemble of impurities,
however it requires both time-reversal symmetry and spatial
homogeneity.  The method described below can be considered a
generalization of the Anderson description which has neither 
of these constraints.

\subsection{The Cooperon}

The problem of the localization of a quantum particle by a random
potential was originally posed by Anderson in 1958 \cite{anderson58},
in the context of understanding metal-insulator transitions, and has
led to a large body of work on the subject of the quantum effects of
disorder in metals.  The field has been summarized in a number of
recent reviews \cite{lee85,sudip86,kramer93,altshuler95}.

One of the theoretical approaches to understanding the quantum effects
of disorder is impurity-averaged perturbation theory
\cite{abrigork,edwards}.  For non-interacting electrons scattered by
rigid impurities, Green's functions may be expanded perturbatively
in the impurity potential.  Averaging over a random ensemble of
impurity potentials yields classes of diagrams which can be organized
by the powers of $k_F^{}\ell$ they contain, where $k_F^{}$ is the
Fermi momentum and $\ell$ is the mean free path.  Diagrams within a
given class may be summed, resulting in a series expansion in
$k_F^{}\ell$ for the desired Green's function.

Impurity-averaged perturbation theory can be used to calculate the
conductivity of an electron gas in the presence of random disorder.
The conductivity is related to the current-current response
function by the Kubo-Greenwood formula, and the impurity diagrams
for the current-current response function can be summed.  It was
recognized early on that a certain class of diagrams, the maximally
crossed ones, yield a significant quantum correction to the classical
conductivity \cite{langer66}.  This estimate for the quantum
correction was used as support for the scaling theory of localization
\cite{gangof4,otherptsums}.

The maximally crossed diagrams for the current-cur\-rent response
function (a particle-hole response function) can be related by
time-reversal to ladder diagrams for a response function in the
particle-particle channel.  The impurity-averaged particle-particle
response function was called the `Cooperon' by Altshuler {\it et al.}
\cite{altsh80,altsh81b}.  It is this response function, the Cooperon,
originally used to study the conductivity of a disordered electron
gas, which appears in the random matrix description of type-II
superconductors described below.

\bs

\def\om{\omega}

\section{Random Matrix Model}

The starting point for an electronic description of type-II
superconductors is the Hamiltonian \cite{gorkov59}
\begin{equation}
 {\cal H}=\int\! d{\bf r} \,c^\dagger_{{\bf r}\,\sigma} {\cal H}_0({\bf r}) 
 c^{\vphantom{\dagger}}_{{\bf r}\,\sigma} - 
{1\over 2}\,V_0 \,\Omega \int\! d{\bf r}\, c^\dagger_{{\bf r}\sigma}
 c^\dagger_{{\bf r}\sigma'} c^{}_{{\bf r}\sigma'} c^{}_{{\bf r}\sigma} .
\label{startingh}
\end{equation}
Here $\Omega$ is the system volume, $V_0>0$ represents an attractive
short-ranged interaction, the creation operators satisfy
\bea
\nonumber
\{c^{}_{{\bf r}\sigma}, c^\dagger_{{\bf r'}\sigma'}\} &=&
\delta_{\sigma\sigma'}\;\delta^3({\bf r}-{\bf r'})\\
\nonumber
\{c^{}_{{\bf r}\sigma}, c_{{\bf r'}\sigma'}\} &=&
\{c^\dagger_{{\bf r}\sigma}, c^\dagger_{{\bf r'}\sigma'}\} =
0\ ,
\eea
and the bare Hamiltonian ${\cal H}_0$ is 
\be
{\cal H}_0({\bf r}) \ \equiv\ {1\over 2m}\;
\left( i{\bf\nabla} - {e\over c}{\bf A(r)}\right)^2+U(r)-E_F^{}\ ,
\label{barehdef}
\ee
with $U(r)$ the potential due to a static distribution of impurities.
The creation operators in the interaction term in
Eq. (\ref{startingh}) are understood to be superpositions of operators
which create states only near the Fermi surface, that is, $c_{\bf
r}=\int \! d{\bf p} \; \exp(i{\bf p}\cdot{\bf r})\, c_{\bf p}\,
\theta_{\bf p}$, where the theta-function limits the energy to be
within a Debye frequency of the Fermi energy: $\theta_{\bf
p}\equiv\theta(\omega_D^{}-|\varepsilon_{\bf p}-E_F^{}|)$.  In the
weak-coupling limit, where the range of the attractive potential is
much smaller than the extent of a Cooper pair ($v_F^{}/\omega_D\ll
v_F^{}/\Delta_0$), the interaction term may be considered local
\cite{agd}.

The variational approach to the Hamiltonian in Eq. (\ref{startingh})
is defined by
\begin{equation}
\label{hprime}
 {\cal H}' = \int \!d{\bf r} \; \, \Psi_{\bf r}^\dagger \;
 \left[ \begin{array}{cc} {\cal H}_0({\bf r}) & \Phi({\bf r}) 
 \\ \Phi^\ast({\bf r})  & -{\cal H}_0^\ast({\bf r}) \end{array} \right]
\; \Psi_{\bf r}^{} \ ,
\end{equation}
where
\[
\Psi_{\bf r}^\dagger \equiv \big[ \, c^\dagger_{{\bf r}\uparrow} \;
c^{}_{{\bf r}\downarrow} \, \big]\ .
\]
The order parameter $\Phi({\bf r})$ is determined self-consistently
from the equation 
\begin{equation}
\label{selfcons}
 \Phi({\bf r}) = -V_0\,\Omega\,
 \langle c^{\vphantom{\dagger}}_{{\bf r}\uparrow} 
c^{\vphantom{\dagger}}_{{\bf r}\downarrow}  \rangle \ .
\end{equation}
For a pure system ($U=0$) with no magnetic field, plane waves
diagonalize the bare Hamiltonian ${\cal H}_0$.  If the order parameter
is assumed to be a constant, $\Phi({\bf r})=\Delta_0$,
then Eq. (\ref{hprime}) separates into $2\times 2$ matrices.  This
yields the standard BCS results: the spectrum is given by $E_{\bf
k}=\sqrt{\varepsilon_{\bf k}^2 + \Delta_0^2}$ , where
$\varepsilon_{\bf k} = k^2/2m - E_F^{}$, and Eq. (\ref{selfcons})
becomes the BCS gap equation $\rho_0 V_0 \int d\varepsilon_{\bf
k}/2E_k=1$, with $\rho_0$ the density of states at the Fermi level.

In the case of a system with impurities ($U\ne 0$), but with no
magnetic field, a similar result holds, as shown by Anderson
\cite{anderson59}.  If we write the eigenstates of the bare
Hamiltonian as
\be
{\cal H}_0(H=0)\, \psi_\alpha^0\ =\ \varepsilon_\alpha\,\psi_\alpha^0
\ee
and rotate to a basis defined by these eigenstates,
\be
c_{\alpha\sigma}^\dagger \equiv \int \! d{\bf r}\ \psi_\alpha^0({\bf r})
\; c_{{\bf r}\sigma}^\dagger \ ,
\ee
then the off-diagonal, pairing term in ${\cal H}'$ becomes
\be
\int \! d{\bf r}\ \Phi({\bf r})\, c_{{\bf r}\aup}^{}\,
c_{{\bf r}\adw}^{} \ =\ 
\tilde\Delta \; \sum_{\alpha\beta}\; 
c_{\alpha\aup}^{}\, c_{\beta\adw}^{}\; {\cal A}_{\alpha\beta}^0 \ ,
\ee
where
\be
{\cal A}_{\alpha\beta}^0 \ \equiv \ 
\int\!d{\bf r}\ \psi_\alpha^0({\bf r}) \psi_\beta^0({\bf r})\ .
\label{azerodef}
\ee
Note that we explicitly assume that the order
parameter is spatially homogeneous: $\Phi({\bf r})=\tilde \Delta$.
The time-reversal symmetry of the system (no magnetic
field, no magnetic impurities) implies that ${\cal H}_0=
{\cal H}_0^\ast$ and therefore a basis may be chosen in which
$\psi_\alpha^0=\psi_\alpha^{0\ast}$.  The overlap integral in
Eq.\ (\ref{azerodef}) then yields
\be
{\cal A}_{\alpha\beta}^0 \ =\ \delta_{\alpha\beta}
\ee
The Hamiltonian ${\cal H}'$ again factorizes into $2\times 2$
matrices, the spectrum is again given by $E_\alpha =
\sqrt{\varepsilon_\alpha^2 + \tilde\Delta^2}$ and solving the
self-consistency equation yields the same result as in the impurity
free case (assuming the density of states at the Fermi level is
unchanged): $\tilde\Delta=\Delta_0$.

The random matrix description we will use for type-II superconductors
\cite{bahcall96} is a generalization of the Anderson description in
the sense that we start by using exact eigenstates of the bare
Hamiltonian ${\cal H}_0$ as a basis.  This includes whatever vortex or
impurity distribution is present.  The eigenstates are
\be
{\cal H}_0\, \psi_\alpha \ =\  \varepsilon_\alpha\,\psi_\alpha \ ,
\ee
and we rotate to the basis
\be
c_{\alpha\sigma}^{} \ \equiv\  \int \! d{\bf r}\ \psi_\alpha({\bf r})
\; c_{{\bf r}\sigma}^{} \ .
\label{newbasis}
\ee
There are two important differences from the zero magnetic field case.
First, the order parameter will no longer be spatially homogeneous;
the phase winds rapidly throughout the superconductor (by $2\pi$ near
the center of each vortex) and the magnitude vanishes at the center of
each vortex.  In order to extract a typical measure of the magnitude
of the superconductivity, we define
\be
\Phi({\bf r}) \ \equiv\ \phi\; \chi({\bf r})\ ,
\label{chidef}
\ee
where $\chi({\bf r})$ is normalized so that $\int |\chi({\bf
r})|^2\,d{\bf r} = \Omega$.  Hence $\phi^2$ is the spatial average of
the magnitude of the order parameter.  Second, we no longer have
time-reversal symmetry, because of the applied magnetic field.  We
still define the pairing matrix as
\begin{equation}
{\cal A}_{\alpha\beta} \ \equiv \ \int\!\!d{\bf r} \  \chi({\bf
r})\,\psi_\alpha^{}({\bf r})\,\psi_\beta^{}({\bf r}) \ ,
\label{adef}
\end{equation}
but we no longer have ${\cal A}_{\alpha\beta}=\delta_{\alpha\beta}$.  
Specifically, writing
\begin{equation}
{\cal A}_{\alpha\beta} \ =\  g_{\alpha\beta} \ h^{1/2}(\ve_\alpha
-\ve_\beta)\ ,
\label{hdef}
\end{equation}
then in zero magnetic field we have $h(\ve_\alpha - \ve_\beta) =
\delta_{\alpha\beta}$.  In the next section we will show that for a
randomly disordered electron gas in a magnetic field, the
$\delta$-function distribution for $h(\ve)$ broadens by an amount 
proportional to the magnetic field.

The motivation behind the random matrix model is that
$g_{\alpha\beta}$ is a material-specific, rapidly varying complex
function, the details of which should not affect the density of states
averaged over a random ensemble of disordered superconductors.  The
average density of states is sensitive to the overall structure of the
pairing matrix elements -- the broadening of the zero-field
$\delta$-function distribution of $h(\ve)$ -- but not to the rapidly
varying $O(1)$ complex numbers which multiply it.

The basic assumption of the random matrix model is therefore that 
average properties of an ensemble of superconductors can be modeled
by choosing complex numbers $g_{\alpha\beta}$ from a random,
uncorrelated distribution:
\be
\Big[ g_{\alpha\beta}^{}\,g_{\alpha'\beta'}^{\ast} \Big]_{\rm av}
\ = \  \delta_{\alpha\alpha'}\; \delta_{\beta\beta'} \ ,
\ee
where the brackets denote an average over the distribution: $[X]_{\rm
av}\equiv (1/N_\mu) \sum_{\mu=1}^{N_\mu} X^{(\mu)}$.  The particular
choice of symmetry for $g_{\alpha\beta}$, whether real symmetric
(GOE), Hermitian (GUE), or complex symmetric, will affect the
eigenvalue spacing distribution but will not affect the density of
states in the large matrix size limit.

\def\wt{\widetilde}

Using the above definitions for $\Phi({\bf r})$ and ${\cal A}$, and
rotating to the basis of bare eigenstates defined by
Eq. (\ref{newbasis}), the Hamiltonian in Eq. (\ref{hprime}) becomes
\begin{equation}
\label{newhprime}
{\cal H'}\  =\ 
\wt\Psi^\dagger \;
\left[ \begin{array}{cc} E_0
  & \phi \, {\cal A} \\
  \phi \, {\cal A}^\dagger & -E_0 \end{array}  \right]
\; \wt\Psi^{}\ ,
\end{equation}
where $E_0={\rm diag}( \ldots \ve_\alpha \ldots)$ is the diagonal
matrix of the eigenvalues of ${\cal H}_0$, and $\wt\Psi^\dagger=
[\ldots c_{\alpha\uparrow}^\dagger\ldots\ \ldots
c_{\beta\downarrow}\ldots]$.  The assumption of uncorrelated random
numbers for $g_{\alpha\beta}$ allows us to determine exactly, in the
large matrix limit, the Green's functions of ${\cal H'}$
\cite{bahcall96}.

\section{The Pairing Matrix and the Cooperon}

The random matrix assumption for $g_{\alpha\beta}$ lets us write down
integral equations for the Green's functions of the Hamiltonian ${\cal
H}'$ for any value of $\phi$, the overall strength of the pairing, and
$h(\omega)$, the average pairing amplitude defined in Eq.\
(\ref{hdef}).  This will only be useful in comparing with experiment
if we have some prediction for, or theoretical understanding of, the
pairing amplitude $h(\omega)$.  For a diffusive system, and a
disordered vortex lattice, there is a simple prediction.

We can rewrite Eq.\ (\ref{hdef}) to express $h(\omega)$ in terms of
the average squared matrix element as a function of energy difference:
\be
h(\omega) = {1 \over N \rho_0}\ \Big[ \;\sum_{\alpha\beta} \;
|{\cal A}_{\alpha\beta}|^2\; \delta(\om-
\ve_\alpha+\ve_\beta) \;\Big]_{\rm av} \ ,
\label{homegadef}
\ee
where the brackets denote averaging over the random ensemble.  The
density of levels of the normal metal, $\rho_0\equiv \rho(E_F^{})$, is
assumed to be independent of energy in the narrow range about the
Fermi energy relevant to superconductivity.  The total number of
levels that are being paired in ${\cal H}'$, which determines the size
$N$ of the matrix, is the level density times the maximum pairing
energy.  This cutoff is usually taken to be the Debye energy, so
$N=\rho_0\omega_D$.  The normalization factor $1/N\rho_0$ in 
Eq.\ (\ref{homegadef}) follows from Eq.\ (\ref{hdef}) and the continuous
limit for the energy levels: $\sum_\alpha \to \int \!
d\varepsilon_\alpha \rho(\varepsilon_\alpha)$.

Note also that we will always be interested in the limit where
$N\to\infty$, because $\omega_D$ is much larger than the level
spacing (as well as the pairing strength $\phi$ and the
bare BCS gap $\Delta_0$).  In order for the Hamiltonian ${\cal H}'$ in
Eq. (\ref{newhprime}) to be interesting in that limit, the matrix
elements ${\cal A}_{\alpha\beta}$ must be $O(1/\sqrt{N})$.  (This is
easy to see with perturbation theory: the shift in energy of some
level $\ve_\alpha$ is $\delta\ve_\alpha= \phi^2 \sum_\beta |{\cal
A}_{\alpha\beta}|^2 / (\varepsilon_\alpha + \varepsilon_\beta)$.  The
$N$ terms in the sum must be canceled by a $1/N$ factor from $|{\cal
A}_{\alpha\beta}|^2$.)

Inserting Eq. (\ref{adef}) for ${\cal A}_{\alpha\beta}$ into 
Eq. (\ref{homegadef}) yields
\begin{eqnarray}
\nonumber
h(\omega) = {1\over N\rho_0}\, \Big[ \; && \sum_{\alpha\beta} 
\int\!\! d{\bf r}\, d{\bf r'} \ \psi_\alpha({\bf r}) \psi_\beta({\bf
r}) \psi_\alpha^\ast({\bf r'}) \psi_\beta^\ast({\bf r'}) \\ 
&& \times\ \chi({\bf r}) \chi^\ast({\bf r'}) \ 
\delta(\om-\ve_\alpha+\ve_\beta) \;
\Big]_{\rm av}\ .
\label{hone}
\end{eqnarray}
We can insert a factor of $1=\int dE \;\delta(E-\ve_\alpha)$ and write
this as
\bea
\nonumber
h(\omega) \ =\ {(2\pi)^{-2} \over N \rho_0}&& \, \int\!\! d{\bf r}\,
d{\bf r'}\,dE \ \Big[ \; \delta G_0({\bf r},{\bf r'},E) \\
\times \; && \delta G_0({\bf r},{\bf r'},E+\omega) \
\chi({\bf r}) \chi^\ast({\bf r'}) \;\Big]_{\rm av} \ , 
\eea
where the bare single particle Green's functions are
\bea
\nonumber
\delta G_0 &\equiv& G_0^+ - G_0^- \\
\nonumber
G_0^{\pm}({\bf r},{\bf r'},E) &\equiv&
\sum_\alpha {\psi_\alpha({\bf r}) \psi_\alpha^\ast({\bf r'})
\over E-\ve_\alpha\pm i0^+} \ .
\eea

To make further progress, we will need to make a key assumption: that
there is, on average, no correlation between the bare single particle
Green's functions $G_0^{\pm}$ and the order parameter product $\chi({\bf
r}) \chi^\ast({\bf r'})$ appearing above:
\bea
\Big[ G_0^{\pm} G_0^{\pm} \, \chi \chi^\ast \Big]_{\rm av} =
\Big[ G_0^{\pm} G_0^{\pm} \Big]_{\rm av} \, 
\Big[ \chi \chi^\ast \Big]_{\rm av} \ .
\eea
The wavefunctions which enter $G_0$ are the eigenstates of the bare
Hamiltonian ${\cal H}_0$, not the eigenstates of the full
superconducting Hamiltonian.  Hence the assumption is that the product
of the bare Green's functions is uncorrelated, on average, with the
order parameter, that is, the positions of the vortices.  Below, we
will only be interested in the case in which the magnetic field is
nearly uniform within the superconductor, $H\gg H_{c1}$.  In this
limit, since the magnetic field in ${\cal H}_0$ is uniform, the
positions of the vortices affect $G_0$ only through a gauge-dependent
phase.  The quantity we are interested in, $h(\omega)$, is
gauge-invariant.  Therefore, with the gauge-invariance of the product
of the two terms properly maintained, and for $H\gg H_{c1}$, there are
no additional correlations between the bare Green's functions and the
order parameter.

With the average over the order parameter product written as
\be
\Big[ \chi({\bf r})
\chi^\ast({\bf r'}) \Big]_{\rm av} \ \equiv \ g({\bf r},{\bf r'})\ ,
\label{gdef}
\ee
the expression for $h(\om)$ then becomes
\begin{eqnarray}
\nonumber
h(\omega) = {(2\pi)^{-2}\over N\rho_0}\  {\rm Re} \,&&\int\!\! d{\bf r}\,
d{\bf r'}\  \Big[ \int\!\! dE\; 
G_0^+({\bf r},{\bf r'},E) \\
\label{hgg}
&& \times \ G_0^-({\bf r},{\bf r'},E+\om) \Big]_{\rm av} \; 
g({\bf r},{\bf r'})\ ,
\end{eqnarray}
where we have used the identity
\bea
\nonumber
\Big( G_0^+({\bf r},{\bf r'},E)\, &&
G_0^-({\bf r},{\bf r'},{E+\omega}) \Big)^\ast = \\
&&G_0^-({\bf r'},{\bf r},E)\,
G_0^+({\bf r'},{\bf r},{E+\omega})  
\eea
and we omit the terms involving $G^-G^-$ and $G^+ G^+$ because they do
not contribute.


As mentioned in the introduction, in weak localization theory quantum
corrections to classical transport are calculated by studying 
averaged values of response functions of non-interacting electrons in a
random disorder potential.  The quantity in brackets in
Eq. (\ref{hgg}) is one such disorder-averaged response function: 
the Cooperon \cite{altsh80,altsh81b}.

The order of ${\bf r}$ and ${\bf r'}$ in Eq. (\ref{hgg}) is important.
If the single particle Green's functions appeared in the form
$G_0({\bf r},{\bf r'},E)\, G_0({\bf r'},{\bf r},E+\om)$, the quantity
in brackets would be just the real-space Fourier transform of the
usual density-density response function.  
It is because ${\cal A}_{\alpha\beta}$ is a matrix element for
electrons in a Cooper pair (the overlap between $\psi_\alpha$ and
$\psi_\beta$ rather than $\psi_\alpha^\ast$ and $\psi_\beta$),
that the order of ${\bf r}$ and ${\bf r'}$ is reversed, and the
bracketed quantity in Eq. (\ref{hgg}) is the Cooperon rather than the
diffuson.

The Cooperon response function $C({\bf r},{\bf r'},\omega)$ is defined
as
\be
C({\bf r},{\bf r'},\omega) =
\Big[ \int\!\! dE\; 
G_0^+({\bf r},{\bf r'},E)\, 
G_0^-({\bf r},{\bf r'},E+\om) \Big]_{\rm av} \ .
\ee
This response function, in the limit in which the mean free path is
much smaller than the magnetic length but much larger than the Fermi
wavelength ($\lambda_F\ll \ell \ll \ell_H^{}
\equiv \sqrt{\hbar c/e H}$), obeys a
diffusive equation of motion \cite{altsh80}
\be
\Big[ -i\omega + D\left(i\nabla+{2e\over \hbar c} {\bf A}\right)^2\Big]
C({\bf r},{\bf r'},\omega)= \delta({\bf r}-{\bf r'})\ .
\ee
The diffusion constant $D=v_F^{}\ell/d$, where $d$ is the
space dimensionality.  This equation defines the Green's function for
the Schroedinger equation of a single particle of charge $2e$ in a
magnetic field.  The solution is therefore
\bea
\nonumber
C({\bf r},{\bf r'},\omega)\; = &&\\
\sum_{nk}\int\!\! {dk_z\over 2\pi} && \ {\varphi_{nk}(x,y) 
\varphi^\ast_{nk}(x',y')\ e^{ik_z(z-z')}
\over -i\omega + (4D/\ell_H^2)(n+1/2) + D k_z^2}\ ,
\label{clls}
\eea
where $\varphi_{nk}(x,y)$ are normalized 2D Landau level wavefunctions 
and we assume a uniform magnetic field $H$ (valid for $H\gg H_{c1}$).

This solution is not sufficient to evaluate $h(\omega)$; we also need
to know the order parameter correlation, $g({\bf r},{\bf r'})$, defined
by Eqs. (\ref{chidef}) and (\ref{gdef}).  The phase of the order
parameter winds by $2\pi$ around each vortex.  In a superconductor
with a perfectly periodic vortex lattice we expect $\Phi({\bf
r})\Phi^\ast ({\bf r'})$ to reflect the long-range ordered structure
of the vortex lattice.  Averaged over an ensemble of disordered
materials, however, the phase will disorder on the length scale set by
the inter-vortex spacing $\ell_H^{}$.  Therefore we will assume that
the averaged correlation function has the form
\be
g({\bf r},{\bf r'}) \ = \ e^{-|{\bf r}-{\bf r'}|^2/\ell_H^2}\  
e^{i\theta_{\rm gauge}({\bf r},{\bf r'})}\ . 
\label{ggaussian}
\ee
The gauge-dependent phase factor is required to keep the
product $C({\bf r},{\bf r'},\omega)g({\bf r},{\bf r'})$
gauge-invariant.  In the Landau gauge ${\bf A}=H x \hat {\bf y}$, we
have $\theta_{\rm gauge}({\bf r},{\bf r'})= (x+x')(y-y')/2\ell_H^2$.

The form of $|g({\bf r},{\bf r'})|$ is an additional assumption that
is necessary for evaluating $h(\omega)$.  It is not essential that
this form be Gaussian.  However, the Gaussian form is natural for a
random system, and also has the property that the result for
$h(\omega)$ is particularly simple.

We now substitute the expressions for $C({\bf r},{\bf r'},\omega)$ and
$g({\bf r},{\bf r'})$, Eqs. (\ref{clls}) and (\ref{ggaussian}), into
Eq. (\ref{hgg}).  The integral over $z-z'$ sets $k_z=0$ in the
denominator.  The Gaussian form for $g({\bf r},{\bf r'})$ can be
written as $|g| = \sqrt{\pi}\, h_0((x-x')/\ell_H^{})
h_0((y-y') / \ell_H^{})$, where $h_n(u)$ are the normalized
harmonic oscillator eigenstates (Hermite functions).  The 2D Landau
level wavefunctions in Landau gauge have the form $\varphi_{nk}(x,y) =
A\, \exp(ikx)\, h_n(k\ell_H^{}/\sqrt{2} + \sqrt{2}y/\ell_H^{})$.
Rewriting the product in Eq.\ (\ref{clls}) using the identity $h_n(a)
h_n(b) = \sum_j C_{nn}^j h_{2n-j}((a+b)/\sqrt{2}) h_j((a-b)/\sqrt{2})$
\cite{bahcall95}, we see that the spatial integrals in Eq.\
(\ref{hgg}) and the orthogonality of the Hermite functions set
$n=0$.  The final result is therefore that only the vanishing
$n$ and $k_z$ terms survive in the sums in Eq.\ (\ref{clls}).  This yields
for $h(\omega)$:
\be
\label{hlorentz}
h(\omega) = {1\over 2\pi\rho_0}\; {W \over \omega^2 + W^2}\ ,
\ee
where the width $W$ is
\be
\label{wdef}
W = 2\hbar D/\ell_H^2 = 2eDH/c\ ,
\ee
and we have normalized to $\rho_0 \int\! d\omega\, h(\omega) =1$.

The Lorentzian form for $h(\omega)$, Eq. (\ref{hlorentz}), is what
might be expected on general grounds in a diffusive system where the
time-evolution of pair correlations obeys an exponential decay law.
This point was discussed by de Gennes in the context of the ``ergodic
evolution of the time-reversal operator'' \cite{deGennes}.  de Gennes
considered the problem of evaluating the transition temperature
$T_c(H)$ for a general superconducting system in a magnetic field
described by the variational Hamiltonian in Eq. (\ref{hprime}).  He used
a perturbative expansion in the order parameter $\Phi({\bf r})$, which
is valid near the phase transition, and found that the average over
single-particle wavefunctions which appears in Eq. (\ref{hone}) is the
kernel operator acting on $\Phi({\bf r})$ in the gap equation.  de
Gennes related that average over single-particle wavefunctions to the
time evolution of the time-reversal operator, $K(t)$, for one electron
moving in the potential described by the Hamiltonian ${\cal H}_0$
(Eq. (\ref{barehdef})).  An ``ergodic'' system was defined as one in
which the motion of the electron is diffusive, in the sense that the
time-reversal operator decays exponentially with time:
\be
\lim_{t\to\infty} \langle K^\dagger(0) K(t) \rangle \ = \
e^{-t/\tau} \ .
\ee
For a system defined as ergodic in this sense, $h(\omega)$, which is
the power spectrum of the operator $K(t)$, has the Lorentzian form
given in Eq. (\ref{hlorentz}), with $W=\tau^{-1}$.

The interesting result here is that the average over the
single-particle wavefunctions in Eq. (\ref{hone}) appears not only in
the equation for $T_c(H)$, but also in the random matrix description
of the pairing Hamiltonian.  The functional form of $h(\omega)$ and
the random matrix technique allow us to go beyond perturbation theory
and therefore beyond calculating $T_c(H)$.  At any field for which the
random matrix element hypothesis is valid, we can obtain the full
Green's functions non-perturbatively.

We note also that the connection with ``ergodic superconductors''
suggests that the random matrix description may apply, in addition to
the bulk disordered systems considered here, to other systems
discussed by de Gennes {\it et al.}  as being ergodic, for example,
thin films and superconducting nanoparticles
\cite{deGennes,dgtinkham}.

\section{Upper critical field}

Using Eqs.\ (\ref{hlorentz}) and (\ref{wdef}) we may derive a relation
between the upper critical field $H_{c2}$ and the diffusion constant
$D$.  This will allow us to test whether the formalism described in
the previous section yields a result consistent with the prediction of
the semiclassical theory of type-II superconductors.  That theory is
based on the perturbative solution of the Gor'kov equations near
$H_{c2}$ using semiclassical Green's functions to include the effects
of the magnetic field.

The difference in energy between the normal metal ground state of the
Hamiltonian in Eq. (\ref{hprime}), given by $\phi=0$, and the
superconducting ground state, is, to second order in $\phi$:
\be
  \Delta E = \phi^2 \left[ \;{1\over V_0} - \sum_{\alpha\beta} {|{\cal
A}_{\alpha\beta}|^2\over |\varepsilon_\alpha+\varepsilon_\beta|}
\;\theta(\ve_\alpha\ve_\beta) \right] + O(\phi^4)\ .
\label{secorder}
\ee
The $|{\cal A}_{\alpha\beta}|^2$ term is second order perturbation
theory for the Hamiltonian Eq.\ (\ref{hprime}), where $\theta(x)$ is
the step function (for non-zero temperatures, $\theta
(\varepsilon_\alpha \varepsilon_\beta)$ is replaced by
$[\tanh(\varepsilon_\alpha/2T)+\tanh(\varepsilon_\beta/2T)]/2$) and
the upper cutoffs for the energies are $|\varepsilon_\alpha|<\om_D$,
$|\varepsilon_\beta|<\om_D$.  The $1/ V_0$ term is the usual
Hartree-Fock self-consistency contribution and follows from $\langle \int
d{\bf r}\; \Phi({\bf r}) c_{{\bf r}\uparrow} c_{{\bf r}\downarrow}
\rangle$ and $\Phi({\bf r}) = -\Omega V_0 \langle c_{{\bf r}\uparrow}
c_{{\bf r}\downarrow} \rangle$.  We must ignore small non-perturbative
effects \cite{bahcallssc} (the breakdown of second-order perturbation
theory), in order to reproduce the semiclassical limit.

The upper critical field is the field $H_{c2}$ at which the quantity
in brackets in Eq. (\ref{secorder}) vanishes.   At lower fields, the
coefficient of $\phi^2$ is negative and $\phi>0$ minimizes the energy;
at higher fields, the coefficient of $\phi^2$ is positive and $\phi=0$
minimizes the energy.

Averaging over the ensemble amounts to replacing $|{\cal
A}_{\alpha\beta}|^2$ in Eq.\ (\ref{secorder}) by its average value
$h(\varepsilon_\alpha-\varepsilon_\beta)$.  Taking the continuum limit
for the energies, inserting Eq. (\ref{hlorentz}) for $h(\nu)$, and
using the definition of the BCS coupling constant, $1/g\equiv 1/\rho_0
V_0$, yields that the $O(\phi^2)$ term in Eq.\ (\ref{secorder}) vanishes
when
\be
{1\over g}\ =\ {2\over\pi}\; \int_0^{\omega_D}\!\!
\int_0^{\omega_D} \,{d\varepsilon_\alpha\,d\varepsilon_\beta\over
\varepsilon_\alpha+\varepsilon_\beta}\  
{ W \over (\varepsilon_\alpha-\varepsilon_\beta)^2+W^2}\ .
\ee
Performing the integrals yields
\be
{1\over g} =  \log\left({2\om_D\over W}\right) \; 
\left[ 1+ O\left({W\over\om_D}\right) \right] \ .
\ee
In the weak coupling limit $W/\om_D\to 0$. 
Subtracting the zero-field BCS result,
\be
{1\over g} =  \log\left({2\om_D\over \Delta_0}\right) \ ,
\ee
where $\Delta_0$ is the zero-field BCS gap, we obtain the equation
which determines the upper critical field:
\be
W(H_{c2}) = \Delta_0\ .
\ee
Inserting Eq. (\ref{wdef}) for $W(H)$ yields
\be 
\label{numerhc2}
H_{c2} = {3\over 2\pi^2} \, {\Phi_0\over \ell\,\xi_0}\ ,
\ee
where $\xi_0=\hbar v_F/\pi \Delta_0$ is the zero-field BCS coherence
length and $D=v_F\ell/3$ in three dimensions.

This result may be compared with the result derived with the
semiclassical theory of dirty type-II superconductors.  The
semiclassical approach starts with plane wave electronic eigenstates,
assumes a weak magnetic field, and includes the effects of impurities
through the Abrikosov-Gor'kov impurity averaged perturbation theory
technique.  The resulting Gor'kov equations can be expanded
perturbatively in the order parameter (near $H_{c2}$) and
yield the equation which determines $H_{c2}$
\cite{dhc2maki,dhc2degennes,dhc2helfand}
\be
\ln\left({T_{c0}\over T}\right)= -\psi\left({1\over 2}\right) + 
\psi\left({1\over 2} + {H_{c2}^d\,D\over 2\,\Phi_0\,T}\right) \ ,
\ee
where $\psi(x)$ is the digamma function, $T_{c0}$ is the zero-field
critical temperature, and $H_{c2}^d$ denotes the dirty limit, $\ell\ll
\xi_0$.  As $T\to 0$ this gives $H_{c2}^d = (3/2\pi^2)
(\Phi_0/\ell\xi_0)$, in agreement with Eq. (\ref{numerhc2}).

The agreement with the semiclassical theory shows that the random
matrix description combined with the results from the Cooperon
problem reproduces the perturbative result of the more conventional
approach to understanding dirty type-II superconductors.

\section{Connection with Abrikosov-Gor'kov Theory, and the $H\to 0$
Limit}

As pointed out in Ref. \cite{bahcall96}, in the weak field limit, for
which the T-breaking scale $W\ll \Delta_0$, the random matrix model
described in the previous sections does {\it not} yield the density of
states the zero-field BCS theory.  What happens in this limit is that
the central assumption of the model, that the Hamiltonian of the
superconductor can be well described by a large matrix with random
pairing elements, breaks down.  For small $W/\Delta_0$ there exists a
special basis, discussed below, in which the Hamiltonian has a simpler
form.  The existence of a special basis means that correlations have
developed which are not obvious in the original basis.  (As a simple
analogy, consider an electron in a cylindrically symmetric potential.
Random matrix elements in a plane wave basis would be a poor assumption.)

To understand what happens to the random matrix model of the
previous sections in the weak field limit, we first write down
two additional possible random matrix models for superconductors with
T-breaking.  The first of these turns out to be a random matrix
formulation of the Abrikosov-Gor'kov theory.   The second of these we
will argue describes the weak field limit, $H\ll H_{c2}$, of a
superconductor in a magnetic field.

The theory described in the previous sections we denote by
$H_I^{}$:
\begin{equation}
 \label{H_I}
 \begin{array}{rcl}
 H_I^{} &=& \left[ \begin{array}{cc} E_0
                  & \phi \, {\cal A} \\
            \phi\, {\cal A}^\dagger & -E_0 \end{array}  \right]\ ,
 \end{array}
\end{equation}
with ${\cal A}$ satisfying 
\[
 \big[ {\cal A}_{ij}^{} {\cal A}_{kl}^\ast \big]_{\rm av} \ =\  
 \delta_{ik}\; \delta_{jl}\; h(\varepsilon_i-\varepsilon_j)\ .
\]
As before, $E_0={\rm diag}( \ldots \ve_\alpha \ldots)$ is a diagonal matrix
of uniformly distributed eigenvalues.

A different possible random matrix model occurs if we start with BCS
and add a random matrix to the diagonal components, in a way which 
breaks the time-reversal symmetry:
\begin{equation}
\label{H_II}
\begin{array}{rcl}
H_{II}^{} &=& 
\left[ \begin{array}{cc} E_0  & \Delta \\
   \Delta & -E_0 \end{array}  \right] + \alpha
\left[ \begin{array}{cc} M  & 0 \\ 0 & M \end{array}  \right] \ ,
\end{array}
\end{equation}
with $M$ satisfying
\[
 \big[ M_{ij}^{} M_{kl}^\ast \big]_{\rm av}\ =\ \delta_{ik}\; \delta_{jl}\
.  \]
$E_0$ is a diagonal matrix as in the previous case, $\Delta$ is
proportional to the identity matrix: $\Delta=\Delta_0\delta_{ij}$,
$\alpha$ is a parameter which measures the amount of T-breaking, and
$M$ is a random Hermitian matrix with {\it no} weighting factor
$h(\varepsilon_i- \varepsilon_j)$ in it.  The T-breaking in this model
is represented by the same sign for $M$ in the upper and lower
diagonal quadrants.  Since these quadrants correspond to the energies
for up-spin electrons and down-spin holes, this indicates a potential
which has opposite sign for up and down spins.  In other words, this
model appears to correspond to random magnetic impurities.  In fact,
the equations for the Green's functions of this random matrix model
are the same as the Abrikosov-Gor'kov equations for a
superconductor in the presence of random magnetic impurities.  This
random matrix model is a reformulation of the AG theory \cite{bkl}.

The final random matrix model we consider is
\begin{equation}
\label{H_III}
\begin{array}{rcl}
H_{III}^{} &=& 
\left[ \begin{array}{cc} E_0  & \Delta \\
   \Delta & -E_0 \end{array}  \right] + \alpha_1
\left[ \begin{array}{cc} M^{(1)}  & 0 \\ 0 & M^{(1)} \end{array} \right] 
\\[4\jot]
&&\qquad +\ \alpha_2
\left[ \begin{array}{cc} 0 &  M^{(2)} \\  M^{(2)} & 0 \end{array} \right] \ ,
\end{array}
\end{equation}
with the matrices $M^{(a)}$ satisfying
\[
  \big[ M_{ij}^{(a)} M_{kl}^{(b)\ast} 
  \big]_{\rm av}\ =\ \delta_{ab}\; \delta_{ik}\; \delta_{jl} \ .
\]
This differs from $H_{II}^{}$ in that a random matrix
appears in both the diagonal and off-diagonal quadrants.

How do we best describe a superconductor in the weak field limit,
$H\ll H_{c2}$?  The order parameter in a superconductor may be written
as
\begin{equation}
\Phi({\bf r})\ = \ f({\bf r})\; e^{i\theta({\bf r})}\ ,
\end{equation}
where $f({\bf r})$ is real and $\theta({\bf r})$ is a position
dependent phase that winds by $2\pi$ around the center of each vortex.
For a given material, with some order parameter phase realization
$\theta({\bf r})$, we may transform the standard creation operators
to a new basis defined by
\begin{equation}
c_{\bf r}\ \to\ \tilde c_{\bf r} \equiv c_{\bf r}\, e^{i\theta({\bf r})/2}\ .
\end{equation}
In this basis the pairing interaction takes the form
\begin{equation}
\Phi({\bf r})\, c_{\bf r\uparrow}^\dagger \, c_{\bf r\downarrow}^\dagger
\ \to\ 
f({\bf r})\, \tilde c_{\bf r\uparrow}^\dagger \, \tilde c_{\bf
r\downarrow}^\dagger\ .
\end{equation}
The effective order parameter in the new basis is purely real.  Note
that although this has sometimes been called a gauge transformation,
it is not: the magnetic field changes.  The vector potential is
$\tilde {\bf A}({\bf r}) = {\bf A}({\bf r}) + \nabla \theta/2e$ so
$\int \tilde {\bf A}({\bf r})\cdot d{\bf l} \ne\int {\bf A}({\bf
r})\cdot d{\bf l}$ around a loop enclosing a vortex.  

Physically, we are transforming to a basis in which the magnetic field
is the previous field distribution plus one flux quantum threaded
through the center of each vortex in a direction opposite to that of
the applied field.  This means that far away from an isolated vortex,
the net effective flux seen by an electron is zero rather one flux
quantum.  In the original basis, the vector potential ${\bf A}$ falls
off as $1/r$ for distances greater than a penetration depth away from
a single vortex.  The transformed vector potential, $\tilde {\bf A}$,
follows the superfluid velocity ${\bf v_s}({\bf r})$, and vanishes
exponentially for distances greater than a penetration depth.
  
The matrix model of the previous sections, $H_I^{}$, uses as a basis
the eigenstates of the full Hamiltonian ${\cal H}_0$, which includes
the magnetic field.  The models described by $H_{II}^{}$ and
$H_{III}^{}$ first apply the above transformation, then use the
the basis defined by the eigenstates of just the T-invariant part of
the Hamiltonian.  This has two important effects when the
magnetic field is weak.

First, weak T-breaking implies that the phase gradient of the order
parameter varies slowly on the scale of a coherence length:
$\xi_0\nabla\theta\ll 2\pi$.  Therefore the term in the bare
Hamiltonian due to the superfluid velocity is small compared to the
energy gap: $(1/m){\bf p}\cdot e\tilde {\bf A}({\bf r})\sim
v_F^{}\nabla\theta \ll \Delta_0$.  Second, weak T-breaking in a
conventional superconductor means that the deviation in the magnitude
of the order parameter from the bare BCS value is small: $\delta
f/\Delta_0 \equiv \int (d{\bf r} / \Omega) |f({\bf r})-
\Delta_0|/\Delta_0 \ll 1$.  In the Ginzburg-Landau description of an
isolated vortex, the order parameter relaxes to the bare value
exponentially, on the scale of the coherence length.  Hence $\delta
f/\Delta_0 \sim \xi_0^2/\ell_H^2 = H/H_{c2}$.  In a more realistic
description, $f({\bf r})$ may have power-law behavior $|f({\bf
r})-\Delta_0| \sim (\xi_0/r)^n$ which yields $\delta f/\Delta_0\sim
(H/H_{c2})^{1/2}$ ($n=1$), or $\delta f/\Delta_0\sim (H/H_{c2})
\ln(H_{c2}/H)$ ($n=2$).  In either case, rotating to a basis of the
eigenstates of the T-invariant part of ${\cal H}_0$ for which the
order parameter is real yields a Hamiltonain which is a small
correction to the BCS theory.

The models $H_{II}^{}$ and $H_{III}^{}$ are defined by the assumption
that the matrix elements of the T-breaking corrections to the BCS
theory are uncorrelated.  That is, in Eqs.\ (\ref{H_II}) and
(\ref{H_III}), the matrices $M$ and $M^{(1)}$ (the superfluid velocity
correction), and $M^{(2)}$ (the order parameter correction), are
assumed to be random.  The difference between the two models is that
in $H_{II}$ the effect of the deviation of the order parameter from
the constant magnitude is ignored, whereas in $H_{III}^{}$ it is
included.  Hence in $H_{II}$ the pairing is diagonal, $\Delta_0
\delta_{ij}$, because $\langle i| T | j \rangle =\delta_{ij}$ ($T$ is
the time reversal operator $T: \psi_i({\bf r}) \to \psi_i^\ast({\bf
r}))$, while in $H_{III}^{}$, the deviation of the order parameter
yields an additional correction: $\langle i| T f({\bf r}) | j \rangle
=\Delta_0\delta_{ij}+\alpha_2 M^{(2)}_{ij}$, where $\alpha_2\sim\delta
f$.

These models describe small corrections to BCS and hence properly
reproduce the BCS result as $H/H_{c2} \to 0$.  The matrix model of the
previous sections, $H_I^{}$, can not be described as a small
correction to BCS.  In the weak field limit, $W/\Delta_0 \sim H/H_{c2}
\to 0$, so $h(\omega)\to \delta(\omega)$.  However, because the
bare level spacing is always assumed to be much less than the
T-breaking scale ($\delta_0 \ll W$), the resulting matrix model still
pairs many levels with equal, random amplitude.  This leads to large
deviations from BCS.  For $W/\Delta_0\to 0$, the density of states has
a node near zero energy, $\rho(E)\sim E$, rather than a gap as in the
BCS spectrum \cite{bahcall96}.

At weak fields, or more generally for weak T-breaking, when the
correction to BCS behavior can be considered small, the most
appropriate random matrix description is that of the third model
listed above: $H_{III}^{}$.  As the T-breaking energy scale increases,
this description must break down.  Although the order parameter can
always be written as the BCS value plus a correction, $f({\bf
r})=\Delta_0 + \delta f({\bf r})$, the assumption that the matrix
element $\langle i|T\,\delta f({\bf r})|j\rangle$ is random is no
longer justifiable when $\delta f/\Delta_0 \sim 1$; $f({\bf r})$ and
$\delta f({\bf r})$ are comparable functions.  At sufficiently high
fields, the only consistent assumption of randomness applies to the entire
pairing matrix, as assumed in $H_I^{}$.  This may called `chaotic'
pairing.  

Figure 1 illustrates schematically the different regimes.

Physically, at low fields, vortices are well separated, bound states
are confined primarily to individual vortices, and the effect of the
superfluid velocity on extended states is small, yielding only a small
correction to the ordered BCS state.  At higher fields, core states
overlap significantly and form energy bands, which interact in a
complicated way with extended states, the superfluid velocity, and
whatever disorder is present.  In the Anderson metal-insulator
transition, delocalization is accompanied by a change from ordered
(Poisson) statistics to chaotic (random matrix) statistics.  In a
type-II superconductor, we are suggesting that the delocalization due
to increasing vortex density (increasing magnetic field) is
accompanied by a transition from a small correction to the BCS theory
to chaotic pairing. This raises interesting possibilities and
questions, discussed in the next section, regarding the nature of such
a transition in a superconductor.

\section{Summary and Conclusions}

When the disorder and magnetic field are sufficiently large in a
type-II superconductor, so that electronic structure calculations and
small deviations from the BCS theory are no longer applicable, the
chaotic pairing picture of the random matrix model may apply.  The
motivation for the model is that the density of states is not
sensitive to details of the rapidly varying $O(1)$ factors in the
matrix elements between electrons in a Cooper pair, but rather to the
average structure of the pairing matrix.

In section III we showed that the average amplitude to pair electrons
is related to the Cooperon response function previously discussed in
the context of weak localization theory.  For a dirty metal this
response function may be calculated and yields a simple form
for the average pairing amplitude, characteristic of a diffusive
system.  

The connection between the average pairing amplitude and diffusion is
related to what was called by de Gennes ``ergodic pairing''.  That
connection was established by de Gennes in the context of evaluating
the critical field of dirty superconductors, $H_{c2}$, using
perturbation theory.  Here we have shown that this connection exists
independent of perturbation theory.  This allows us to evaluate the
density of states for fields far away from the regime in which
perturbation theory is valid.

Finally, we considered the weak field limit of the random matrix
model, $H\ll H_{c2}$.  At sufficiently weak fields the random matrix
picture must break down because the system must approach the ordered
BCS state.  This suggests a crossover between a low-field ordered
state, in which the corrections to BCS theory are small and can be
understood as being due to states localized near individual vortices,
to a high-field chaotic state, in which most of the states are
extended.  A useful analogy is the Anderson metal-insulator
transition.  Instead of varying the electron density (chemical
potential) from a regime of localized states to a regime of
delocalized states, in the superconductor we vary the vortex density
(magnetic field) between a regime of bound states to a regime of
extended states.  In the metal-insulator transition we have a change
in statistics from Poisson to random matrices; in the superconductor
we have a change in the pairing, from BCS-like to chaotic.

The connection between the field-dependence of the pairing in a
superconductor and the nature of the low-energy quasiparticles raises
several interesting questions.  Is there a true localization phase
transition for quasiparticles?  Is there a sharp change between an
ordered regime and a chaotic regime or merely a crossover?  Are there
simple models for which, using techniques similar to those in the
theory of disordered metals, the chaotic pairing picture may be
derived?  What is the interplay between localization, disorder, and
magnetic field for high-T$_c$ superconductors, where the prediction in
zero field for a two-dimensional $d$-wave superconductor is
``universal conductivity'' \cite{plee}?  A further important issue is
to test the validity of the random matrix prediction for the density
of states, for each of the possible regimes, by direct comparison with
tunneling experiments.

This work was supported by a Miller Post-Doctoral Fellowship from the
Miller Institute for Basic Research in Science.  I would like to thank
D.-H.\ Lee for additional support and useful discussions.

 \begin{figure}[b]
 \epsfysize=6cm\epsfbox{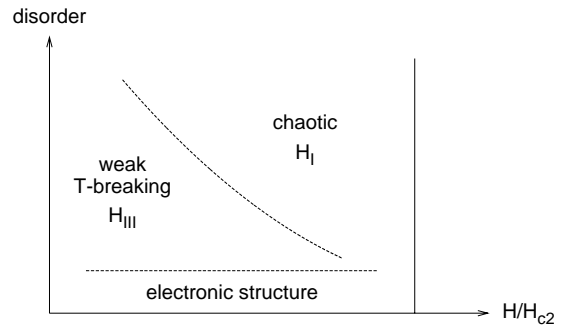}
 \vspace{6pt} \caption{ In the absence of impurities and for an
 ordered lattice of vortices, the quasiparticle spectrum of a type-II
 superconductor may be calculated by numerically solving the
 Bogoliubov-de Gennes equations.  In the presence of random disorder,
 and for weak magnetic fields, the quasiparticle spectrum may be
 calculated as a small correction to the BCS theory, 
 Eq.\  (\protect\ref{H_III}), similar to the Abrikosov-Gor'kov theory.
 For stronger disorder or magnetic fields, the system enters a chaotic
 regime in which the pairing is completely random, 
 Eq.\ (\protect\ref{H_I}).  } \end{figure}


\end{document}